\documentclass[11pt]{article}
\usepackage[T1]{fontenc}
\usepackage[utf8]{inputenc}
\usepackage{amsmath,amssymb,amsthm,geometry,hyperref,xcolor}
\title{Toxicity Bounds for Dynamic Liquidation Incentives}
\author{Alexander McFarlane}
\date{\today}

\begin{document}
\maketitle

\begin{abstract}
We derive a slippage-aware toxicity condition for on-chain liquidations executed via a constant-product automated market maker (CP-AMM). For a fixed (constant) liquidation incentive \(i\), the familiar toxicity frontier \( \nu < 1/(1+i) \) tightens to
\(\nu < 1/(1+i)\lambda\) for a liquidity penalty factor, \(\lambda\), that we derive for both the CP-AMM and a generalised form. Using a dynamic health-linked liquidation incentive \(i(h) = i(1 - h)\), we obtain a state-dependent bound and, at the liquidation boundary, a liquidity depth-only condition \( \,v < 1/\lambda\). This reconciles dynamic incentives with the impact of the CP-AMM price and clarifies when dynamic liquidation incentives reduce versus exacerbate spiral risk.
\end{abstract}

\section{Introduction}
Liquidations in on‑chain lending repay debt by selling collateral, typically with a bonus paid to liquidators. When sales route through AMMs, execution moves prices: the sale lowers the pool price and the borrower’s remaining collateral is re‑marked. If the loss on the remainder exceeds the debt reduction, health falls after the step; repeating such steps creates a \emph{toxic liquidation spiral}. Protocols prefer \emph{partial} liquidations to limit slippage, yet in the toxic regime each partial step increases LTV and drives the position toward \emph{full} liquidation or bad debt.

We consider a borrower with collateral value \(c\) (measured in units of the debt asset) and debt \(q\). Let
\[
\ell := \frac{q}{c} \quad \text{(LTV)}, 
\qquad v \in (0,1) \quad \text{(protocol LLTV parameter)}.
\]
for health \( h := v\,\frac{c}{q} = v/\ell \). Upon a small liquidation repaying \(da\) units of debt, the liquidator is entitled to a bonus \(i \ge 0\) (possibly state dependent); a fraction \((1+i)\,da\) of collateral value is seized.

\section{Slippage-aware toxicity}

\subsection{CP-AMM price impact model}
We consider liquidations that are routed through a CP-AMM with reserves \((x,y)\) for (collateral,debt), price \(P=y/x\), and invariant \(xy=k\). The local price impact of the CP-AMM is
\begin{equation}
d(\ln P) \;=\; d(\ln y - \ln x) \;=\; -2\,\frac{dx}{x}
\label{eq:cp-impact}
\end{equation}
The sale \((1+i)\,da\) of \emph{value} in collateral implies \(dx = \frac{(1+i)\,da}{P}\), and hence \(d(\ln P) = -\frac{2(1+i)}{y}\,da\).

Let collateral $c=sP$, with $s$ the remaining units and $P$ the price. The infinitesimal change in the collateral value is $dc = P\,ds + s\,dP$ to the first order, neglecting $dsdP$. The seizure contributes $P\,ds=-(1+i)\,da$, and the price move re-marks the remainder by $s\,dP=c\,d(\ln P)$. Therefore, the remaining collateral is re-marked by
\[
c\,d(\ln P) = -\frac{2c(1+i)}{y}\,da,
\]
and direct seizure removes \((1+i)\,da\), so
\[
dc = - (1+i)\,da \;-\; \frac{2c(1+i)}{y}\,da
= -(1+i)\Bigl(1+\frac{2c}{y}\Bigr)da,
\qquad dq=-da.
\]
Differentiating \(h = v c/q\) yields
\[
dh \;=\; \frac{v}{q}\!\left[dc - \frac{c}{q}\,dq\right]
\;=\; \frac{v}{q}\!\left[-(1+i)\!\left(1+\frac{2c}{y}\right) + \frac{c}{q}\right]\!da
\]
A liquidation is \emph{toxic} (reduces health) iff \(dh<0\), i.e.
\begin{equation}
\frac{c}{q} \;<\; (1+i)\!\left(1+\frac{2c}{y}\right)
\qquad\text{equivalently}\qquad
\ell \;>\; \frac{1}{(1+i)\,\lambda},
\quad \lambda := 1 + 2\,\frac{c}{y}.
\label{eq:constant-bonus}
\end{equation}
In the infinite-liquidity limit \(y\to\infty\) (so \(\lambda\to 1\)), this reduces to the constant incentive frontier \(\ell < 1/(1+i)\) \cite{WCP2022}.

\subsection{Linear price impact model}
Warmuz, Chaudhary and Pinna \cite{WCP2022} propose a linear slippage model to capture execution costs in decentralised liquidations:
\[
s(x) = \gamma + \frac{\sigma}{L}\,x,
\]
where $s(x)$ is the relative price discount on trade size $x$, $\gamma$ is the spread, $\sigma$ is the slippage parameter, and $L$ is a liquidity scale. The execution price is $1-s(x)$ relative to the oracle.

Linearising around small trades yields an effective per-unit price impact
\[
\phi = \frac{\sigma}{L(1-\gamma)}
\]
This is directly analogous to Kyle’s $\lambda$ in market microstructure theory \cite{Kyle1985}, which measures the permanent price impact per unit of order flow. The slippage penalty factor now becomes
\[
\lambda = 1 + \phi c
\]

\section{Removing toxicity}
We choose a linear incentive function linked to health, increasing as health falls,
\[
i \to i(h) \;=\; i\,\bigl(1-h\bigr)
\;=\; i\!\left(1 - \frac{v}{\ell}\right),
\]
capped at the protocol maximum \(i\). Substituting \(i(h)\) into \eqref{eq:constant-bonus} gives
\begin{equation}
\ell \;>\; \frac{1 + i\,v\,\lambda}{(1+i)\,\lambda}.
\label{eq:euler-dutch}
\end{equation}

\paragraph{Boundary condition (model-agnostic).}
At the LLTV boundary we have $\ell = v$ (equivalently $h=1$). Since the linear function satisfies $i(h)=i(1-h)=0$ at $h=1$, substituting $\ell=v$ into \eqref{eq:euler-dutch} removes any dependence on $i$ and yields a depth-only criterion:
\begin{equation}
\, v \;\le\; \frac{1}{\lambda}\,
\label{eq:euler-result}
\end{equation}

\noindent
This statement is \emph{model-agnostic}: it holds for any monotone impact summarised by a penalty factor $\lambda$. For a CP-AMM, $\lambda=1+2c/y$; for the linear (Kyle) model, $\lambda=1+\phi c$. Writing the result in terms of $\lambda$ avoids unnecessary specialisation and makes clear that a greater depth (smaller $\lambda$) relaxes the admissible LLTV $v$.

\section{Discussion and limitations}
Equation \eqref{eq:euler-result} isolates CP-AMM \emph{depth} as the critical determinant of safety at the LLTV boundary: greater depth (larger \(y\)) lowers \(\lambda\) and raises the allowable \(v\). The derivation is local (infinitesimal step, CP-AMM); integrating over large sales or routing across venues is straightforward in principle but model-specific. Nevertheless, the local condition precisely characterises when a liquidation step is health-improving versus health-worsening and reconciles dynamic incentives with price impact.

\end{document}